\documentclass[pra,aps,twocolumn,10pt,superscriptaddress,notitlepage,longbibliography]{revtex4-2}

\usepackage[dvipsnames]{xcolor}
\usepackage[english]{babel} 
\usepackage{graphicx}
\usepackage{float}
\usepackage{braket}
\usepackage{epstopdf}
\usepackage{mathtools}
\usepackage{amsmath}
\usepackage{amssymb}
\usepackage{bbm,bm}
\usepackage[caption=false]{subfig}
\usepackage{pythonhighlight}
\usepackage{hyperref}
\usepackage{float}
\usepackage{bbold}
\usepackage[T1]{fontenc}

\usepackage[markup=underlined]{changes}
\makeatletter
\@namedef{Changes@AuthorColor}{magenta}
\colorlet{Changes@Color}{magenta}
\makeatother
\usepackage{hyperref}
 \hypersetup{
     colorlinks=true,
     linkcolor=blue,
     filecolor=blue,
     citecolor = magenta,      
     urlcolor=red,
     }

\usepackage{braket}
\newcommand{\blue}[1]{{\color{black} #1}}

\usepackage{xargs}
\newcommandx{\greencom}[2][1=]
{\todo[inline, color=green!40,#1]{#2}}
\newcommandx{\bluecom}[2][1=]
{\todo[inline, color=blue!40,#1]{#2}}
\newcommandx{\bluemargin}[2][1=]
{\todo[color=blue!40,#1]{#2}}

\usepackage{letltxmacro}
\LetLtxMacro{\ORIGselectlanguage}{\selectlanguage}
\makeatletter
\DeclareRobustCommand{\selectlanguage}[1]{%
  \@ifundefined{alias@\string#1}
    {\ORIGselectlanguage{#1}}
    {\begingroup\edef\x{\endgroup
       \noexpand\ORIGselectlanguage{\@nameuse{alias@#1}}}\x}%
}
\newcommand{\definelanguagealias}[2]{%
  \@namedef{alias@#1}{#2}%
}

\makeatother

\definelanguagealias{en}{english}
\definelanguagealias{EN}{english}
\definelanguagealias{eng}{english}
\definelanguagealias{de}{ngerman}

\graphicspath{{figs/}}

\begin{document}

\title{Cavity-like strong coupling in macroscopic waveguide QED using three coupled qubits in the deep non-Markovian regime}

\author{Sofia Arranz Regidor}
\email{18sar4@queensu.ca}
\affiliation{Department of Physics,
Engineering Physics and Astronomy, Queen's University, Kingston, Ontario, Canada, K7L 3N6}
\author{Stephen Hughes}
\affiliation{Department of Physics,
Engineering Physics and Astronomy, Queen's University, Kingston, Ontario, Canada, K7L 3N6}
\email{shughes@queensu.ca}

\date{\today}

\begin{abstract} 
We introduce a three qubit waveguide QED system to mimic the cavity-QED
strong coupling regime of a probe qubit embedded in atom-like mirrors, which was realized in recent experiments. We \blue{then} extend this system into the deep non-Markovian regime and demonstrate the profound role that retardation plays on the dressed
resonances, allowing one to significantly improve the polariton lifetimes (by many orders of magnitude), tune the resonances, as well as 
realize  Fano-like resonances and simultaneous coupling to multiple cavity modes. Exact Green function solutions are presented for the spectral resonances, and the \blue{quantum} dynamics are simulated using matrix product states, where we  demonstrate additional 
control of the cavity-QED system using chiral probe qubits.
\blue{We also show example qubit dynamics with both one and two  quantum excitations}.
\end{abstract}

\maketitle

%\blue{\section{Introduction}}
%\label{sec:introduction}
\textit{Introduction.---}Cavity-QED phenomena have been studied both theoretically and experimentally 
for various cavity system (e.g., see Refs.~\cite{PhysRevA.81.042311,PhysRevLett.125.263606,PhysRevLett.89.067901,PhysRevLett.64.2499,Shamailov2010,PhysRevLett.120.093602,PhysRevLett.120.093601,PhysRevA.92.063830,Yoshie2004,Reithmaier2004,PhysRevLett.95.067401,PhysRevLett.76.1800,Schuster2008,Bishop2008,Fink2008,PhysRevLett.104.073904,PhysRevLett.98.117402}), showing a variety of applications and rich quantum physics. These
cavity systems  are 
typically characterized by Markovian decay dynamics, where interactions between system operators are essentially instantaneous. One of the key signatures of
such systems is the ``strong coupling'' regime, where the intrinsic quantum coherence overcomes  system losses.

A fundamentally different dynamic can be realized using 
waveguide QED systems \cite{Hughes2004,Shen:05,Zheng2010,PhysRevA.83.063828,PhysRevA.102.023702,PhysRevB.81.155117,RevModPhys.89.021001,PhysRevLett.121.123601,PhysRevResearch.2.043213,PetrovWaveguideQED}, which allow one to efficiently couple quantum emitters and two level systems acting as qubits, over macroscopic length scales. Here the waveguide modes mediate the photon coupling between the qubits, and the system can have a non-Markovian memory \cite{PhysRevLett.124.043603,Longhi:21,PhysRevA.103.053701,PhysRevResearch.2.043014,Longhi:20}. Coherent feedback systems can thus lead to non-trivial dynamical protocols such as population trapping, and photon transport beyond the usual dipole-dipole interaction regimes \cite{cajitas,PhysRevResearch.3.023030, crowder_quantum_2020}. For example, these waveguide-QED regimes can be experimentally realized in chip-based systems with semiconductor quantum dots and semiconducting circuits \cite{Gu2017,PhysRevLett.120.140404,PhysRevLett.121.123601}. 

\begin{figure}[t]
    \centering
    \includegraphics[width=1\columnwidth]{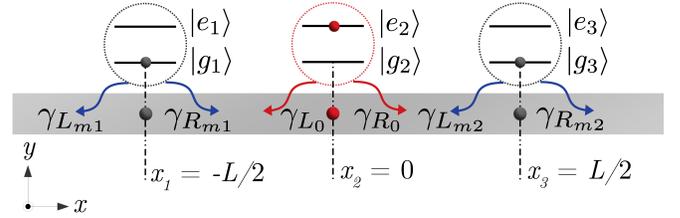}
    \caption{Schematic of three qubits on a waveguide.
    }
    \label{FFT}
\end{figure}

It is interesting to explore when these two disparate systems \blue{(cavities and waveguides)} overlap with each other, in terms of the connecting physics,  and in particular if one can realize familiar cavity-QED phenomena with richer non-Markovian dynamics than what is often assumed with traditional cavities.
To help address this question, 
recently it was experimentally demonstrated how one can achieve cavity-QED phenomena, such as vacuum Rabi oscillations and qubit {\it strong coupling}, by
simply having some of the qubits  act as
``atom-like mirrors''~\cite{Mirhosseini2019,Chang_2012}. 
A schematic representation of this scenario is shown in 
Fig.~\ref{FFT}, where a {\it probe qubit} is situated between two external {\em mirror qubits}.
\blue{Reference~\cite{Mirhosseini2019}}
 showed how the {\em usual} strong coupling regime can be reached by precisely positioning the external qubits, which can behave as a collective entangled state, creating a cavity-like system. This regime is mediated by coherent qubit interactions~\cite{Mirhosseini2019}. This recent experimental achievement \blue{allows one} to manipulate correlated dissipation and coherence of quantum emitter arrays,
with engineerable feedback. 
Strong coupling 
dynamics have
%has 
also been 
theoretically demonstrated for qubits 
at a waveguide mode edge~\cite{PhysRevA.85.043830}.

Motivated by these emerging experiments on tunable quantum networks, here we present a study of qubit-mirrors in {\it the deep non-Markovian regime}, and show how longer time delays between the qubits result in much richer coupling dynamics than traditional cavity-QED, yielding spectrally narrow linewidths and the simultaneous coupling to multiple cavity modes for sufficiently long round trip times. We first employ a classical scattering theory, solving  for the exact Green function to analyze the spectral resonances beyond the Markovian approximation, in a strong coupling regime. We find that the main two Rabi split resonances (from an infinite set) depend on the mirror-qubit delay times, which significantly narrow when the retardation increases, and recover the Markovian limit solution \blue{only} for small delay times. We show explicitly how  a non-Markovian delay time significantly improves the cavity-QED resonances.
We also study this problem from
the viewpoint of  matrix products states (MPS), which allows us to solve the quantum Hamiltonian also in the non-Markovian regime \cite{PhysRevResearch.2.013238} \blue{as well as the nonlinear regime}. We directly show how the lifetimes {\em improve dramatically} for long delay times when the retardation is considered, giving a method to precisely control the coherence times of these novel polariton states. In addition, we investigate the time dynamics for a chiral probe qubit \cite{PhysRevLett.115.153901,Lodahl2017},
\blue{and explore qubit dynamics with one and two initial quantum excitations.}

%\blue{\section{Classical Scattering Theory}}
%\label{sec:classical}
\textit{Classical Scattering Theory.---}Light propagation through an arbitrary dielectric medium can be described in terms of the mode solutions to the Helmholtz equation: 
$\nabla \times \nabla \times \mathbf{f}_{\lambda}(\mathbf{r}) - \frac{\omega_{\lambda}^2}{c^2} \epsilon(\mathbf{r})\mathbf{f}_\lambda(\mathbf{r})=0,$
where $\epsilon(\mathbf{r})$ describes the relative permittivity of the structure and $\mathbf{f}_\lambda(\mathbf{r})$ are generalized field modes with a harmonic $e^{-i\omega t}$ time dependence.  
For the waveguide, 
we consider a photonic nanowire,
supporting lossless waveguide modes
\blue{$\mathbf{f}_{k_\omega}(\mathbf{r})=\sqrt{\frac{1}{L_W}}\mathbf{e}_{k_\omega}(\bm{\rho})e^{i k_\omega x }$}, with $\mathbf{e}_{ k_\omega}(\bm{\rho})$  the mode solution, normalized from \blue{ $\int_{A_{\rm w}} \epsilon(\bm{\rho})\mathbf{e}_{ k_\omega}^*(\bm{\rho})\cdot \mathbf{e}_{k_\omega'}(\bm{\rho})=\delta_{k_\omega, k_\omega'}$}, where $A_{\rm w}$ is the spatial area, and \blue{$L_{W}$} is the length of the waveguide. The waveguide  Green function  is~\cite{PhysRevB.75.205437}
\begin{align}
\mathbf{G}_{\rm w}(\mathbf{r}, \mathbf{r'}, \omega) &=
\frac{i \omega}{2 v_g}\Big[ \Theta(x-x')\mathbf{e}_{k_\omega}(\bm{\rho})
\mathbf{e}^*_{ k_\omega}(\bm{\rho}')e^{i k_\omega (x-x') } \nonumber \\
&+ \Theta(x'-x)\mathbf{e}^*_{k_\omega}(\bm{\rho})\mathbf{e}_{ k_\omega}(\bm{\rho}')e^{i k_\omega (x'-x) } \Big],
\label{eq:Gwg}
\end{align}
where the terms preceded by Heaviside functions correspond to forward and backwards propagating modes, respectively, and $v_g$
is the group velocity at the frequency on interest.
Since the modes are translationally invariant in $x$, then 
$\mathbf{e}_{k_\omega}({\bf r})=\mathbf{e}_{k_\omega}(\bm{\rho})$. The notation can be easily generalized for photonic crystal waveguides~\cite{PhysRevB.75.205437}.

Next, we consider adding in a 
single 
%probe 
qubit treated
at the level of a polarization dipole. The polarizability of the qubit, with resonance energy $\omega_0$, is described through the polarizability tensor
$
{\bm \alpha}_0 = \alpha_0  {\bf n}_d {\bf n}_d^\dagger$ ,
where ${\bf n}_d$ is a unit vector 
describing the polarization direction of the qubit dipole,
and
$\alpha_0 = {A_0\omega^2}/{(\omega_0^2-\omega^2)}$
is the ``bare polarizability'' volume~\cite{RevModPhys.70.447}, i.e., it does not include radiative coupling effects to the environment, and
$A_0 = 2 \omega_p d_0^2/\epsilon_0\hbar$.

The total electric field in the waveguide is ($\omega$ is implicit)
%\begin{align}
$\mathbf{E}(\mathbf{r})=\mathbf{E}^{\rm h}(\mathbf{r})
+ \mathbf{G}(\mathbf{r}, \mathbf{r}_{d})\cdot
{\bm \alpha}_0 \cdot {\bf E}({\bf r}_d)$,
%\label{eq:eprop1}
%\end{align} 
where ${\bm \alpha}_0$ has units of volume, ${\bf r}_d$ is the position of the qubit, \blue{$\mathbf{E}^{\rm h}(\mathbf{r})$ is the homogeneous field solution in the absence of the qubits}. It is also convenient to exploit the Dyson equation, ${\bf G}^{(1)} = {\bf G}^{} +{\bf G}^{} \cdot {\bm \alpha}_0 \cdot {\bf G}^{(1)}$, where the `1' superscript denotes the GF with the addition of one \blue{qubit}, so that 
\begin{equation}
\mathbf{E}(\mathbf{r})=\mathbf{E}^{\rm h}(\mathbf{r})
+ \mathbf{G}^{(1)}(\mathbf{r}, \mathbf{r}_{d})\cdot 
{\bm \alpha}_0 \cdot {\bf E}^{\rm h}({\bf r}_d),
\label{eq:eprop2}
\end{equation}
where 
$\mathbf{G}^{(1)}(\mathbf{r},{\bf r'})
={\mathbf{G}^{}(\mathbf{r},{\bf r'})}/
{(1-\alpha_0 {\bf n}_d^\dagger \cdot {\bf G}_d \cdot {\bf n}_d )}$.
%\begin{equation}
%\mathbf{G}^{(1)}(\mathbf{r},{\bf r'})
%=\frac{\mathbf{G}^{}(\mathbf{r},{\bf r'})}
%{1-\alpha_0 {\bf n}_d^\dagger \cdot {\bf G}_d \cdot {\bf n}_d }.
%\end{equation}
Alternatively, one can introduce
the renormalized polarizability,
${\bm \alpha}^{(1)} = {\bm \alpha}_0 +\ {\bm \alpha}_0 \cdot {\bf G}_d \cdot {\bm \alpha}$, with  ${\bf G}_d \equiv {\bf G}({\bf r}_d,{\bf r}_d)$.  The polarizability, including 
coupling to the medium
%(and allowing for complex dipoles in a Cartesian coordinate system), 
is 
\begin{equation}
{\bm \alpha}^{(1)} =  \frac{A_0 \omega_0^2\, 
{\bf n}_d^\dagger
{\bf n}_d}{\omega_0^2-\omega^2 -i\omega_0\gamma },
\end{equation}
where $\gamma=A_0 \omega_0\, {\bf n}_d^\dagger \cdot {\rm Im} {\bf G}_d 
\cdot {\bf n}_d =\gamma_{\rm L}+\gamma_{\rm R}$, and we assume the qubit is symmetrically
coupled to both forward and backwards modes 
\blue{(unless stated otherwise)}.
It is also useful to define the
normalized scattered field at the qubit,
from $\tilde{E}_{\rm s}^{(1)}({\bf r}_d) = 
{\bm \alpha}_0 \cdot {\bf G}^{(1)}({\bf r}_d,{\bf r}_d) \cdot {\bf n}_d$.
%    = r_1(\omega)$.

Next we consider an injected waveguide mode from the left,
\blue{$\mathbf{E}^{\rm h}(\mathbf{r})=\mathbf{f}_{k_{\rm h}}(\mathbf{r})=\sqrt{\frac{1}{L_W}}
\mathbf{e}_{k_{\rm h}}(\mathbf{r})e^{i k_{\rm h} x }$. }
% For a sufficiently long waveguide, 
The reflection coefficient from one qubit is
% \begin{equation}
% t_{1}(\omega) = \frac{{\bf E}_{\rm t}(\mathbf{r}; x\rightarrow \infty)}{{\bf E}^{\rm h}(\mathbf{r}; x\rightarrow \infty)}
% = 1 + \frac{i\omega_0\gamma}
% {\omega_0^2-\omega^2-i\omega_0\gamma},\\
% \end{equation}
%and
\begin{equation}
r_1(\omega) = \frac{{\bf E}_{\rm r}(\mathbf{r}; x\rightarrow -\infty)}{{{\bf E}^{\rm h}}(\mathbf{r}; x\rightarrow -\infty)}
=  \frac{i\omega_0\gamma e^{i\phi(x_d)}}
%e^{2ik_h x_d}}
{\omega_0^2-\omega^2-i\omega_0
\gamma},
\label{r:1dot}
\end{equation}
\blue{where
% \begin{align}
% \begin{split}
%   &\mathbf{E}_{\rm r}(\mathbf{r}, x\rightarrow -\infty)= \\
%  &\mathbf{G}_{\rm w}(\mathbf{r};x\rightarrow-\infty, \mathbf{r}_{d})\cdot
% {\bm \alpha}^{(1)} \cdot 
% \sqrt{\frac{1}{L}}\mathbf{e}_{k_{\rm h}}(\mathbf{r}_d)e^{i k_{\rm h} x_{d} }, 
% \end{split}
% \label{eq:etl}
% \end{align}
$\mathbf{E}_{\rm r}$ is the reflected field,} and $\phi(x_d)$ is  a positional dependent phase; 
the transmission coefficient can be derived in a similar way.
Thus, we see that the waveguide qubit indeed acts as a mirror,
with a Lorentzian lineshape, whose spectral width is determined by the radiation decay rate. The strategy
for obtaining atom-like mirrors then becomes clear: we can surround a single qubit with two mirror
qubit and expect the system to mimic a Fabry-P\'erot resonator.
To make this clearer, we now add in another
qubit, acting as a second resonant mirror, separated from the first by a distance $L$. 
Assuming identical mirror qubits,
we can use the Dyson equation again to derive 
the total reflection coefficient as
\begin{equation}
\blue{r_{\rm 2dots}}(\omega)=
\frac{r_1(\omega) \left [1+e^{2ikL}+2r_1(\omega) e^{2ikL}\right ]}
{1-r_1^2(\omega) e^{2ikL}},
\end{equation}
which is precisely the solution expected from a 1d cavity with two identical mirrors with a complex reflection coefficient,
$r_1$. \blue{For simplicity, he have neglected the mirror phase terms.}
The {\it mirror} round trip time
is
$ \tau_{\rm RT}
    = {n_g L}/{c}=2\tau$ (with $n_g$ the group index),
which will yield cavity modes with a free spectral range (FSR)
$\Delta\omega_{\rm FSR}\approx 2\pi/\tau_{\rm RT}$.
Note if we make a rotating wave approximation,
\blue{ $\omega_0^2-\omega^2 \approx
2 \omega_0 (\omega_0-\omega)$,} 
then we recover the quantum mechanically derived results~\cite{PhysRevA.95.053807}.

We next add in a probe  (third) qubit, also with $\omega_p=\omega_0$, 
and use the Dyson equation again to derive an explicit solution for the Green function and for the 
scattered fields. At the probe qubit (${\bf r}_p$), we have
\begin{align}
%    \alpha_3^{(3)}(\omega)
    \tilde{E}_{\rm s}^{(3)}({\bf r}_p) &= {\bm \alpha_0} \cdot {\bf G}^{(3)}
     ({\bf r}_p,{\bf r}_p) \cdot {\bf n}_p =\frac{i \omega_p\tilde\gamma_p}{\omega_p^2-\omega^2-i\omega \tilde\gamma_p},
     \label{eq:E3_exact}
    \end{align}
with the modified decay rate
\begin{align}
\tilde\gamma_p =
\gamma_p &\bigg[1+  e^{ikL}r_1(\omega)+
(e^{ikL/2} +  e^{ikL/2} e^{ikL}r_1(\omega)
) \times \nonumber \\
& \ \ \frac{r_1(\omega)\left(e^{ikL/2}+r_1(\omega)e^{ikL/2}e^{ikL}\right)} {1-r_1^2(\omega)e^{2ikL} } \bigg], 
\label{eq:gamma3_exact}
\end{align}
% %
which is an exact solution. It is also straightforward to  derive the scattered field from the mirror qubits, but this solution alone is enough to probe the system resonances.
%

%\blue{\section{Strong Coupling Regime}}
\textit{Strong Coupling Regime.---}
To connect to the familiar cavity-QED strong coupling regime, we can study the
complex poles in Eq.~\eqref{eq:E3_exact}.
Strong coupling is optimally achieved when the modified 
probe decay rate approaches infinity, and 
when \blue{$\omega=\omega_0=\omega_p$}, 
$r_1 \rightarrow -1$. Thus, one needs
to 
satisfy the following phase matching conditions:
%\sh{not sure if we should say that the mirror phase is defined as zero/neglected here, or could also just state that after eq. (4)}
\blue{$e^{i2\omega\tau}\equiv e^{i2kL} =1$
and $e^{i\omega\tau}=-1$}.
Neglecting retardation effects
(Markov approximation),  this is possible when
\blue{$\omega_0 \tau/2    = m\,\pi + \pi/2$},
which was achieved in Ref.~\cite{Mirhosseini2019} by having
%$L_{12}=\lambda/4$
%and 
$L=\lambda/2$ (\blue{$m=0$}), \blue{and the probe qubit at cavity center}.
Using this phase conditions, and a rotating wave approximation, the two main complex roots are:
\begin{equation}
\tilde\omega^{\pm} \approx     \omega_p -i\frac{\gamma_p}{4}
    \pm \frac{1}{2}\sqrt{2\gamma_p \gamma_m} \, ,
\end{equation}
where we assume $\gamma_m \gg \gamma_p$, with $\gamma_m$ the mirror qubit decay rate.
For on-resonance,
\blue{$\omega=\omega_0$}, then
$r_1 \rightarrow -1$, and the term in brackets in
Eq.~\eqref{eq:gamma3_exact} tends to infinity, as does the Purcell effect of an immutable dipole. However, 
%the Purcell
%effect does not work in this regime, since we are not in a weak coupling regime.
in a strong coupling regime, 
with small $\tau$,
the original linewidth is reduced
by 50\% 
%(so this is not the same as optical trapping) 
and there will be a splitting of $\pm g_0$,
with $g_0 =\sqrt{2\gamma_m\gamma_p}/2$~\cite{Mirhosseini2019}.
For longer delay times, the polariton states decrease in energy and their lifetime
increases, an effect that requires retardation in the model.

\begin{figure}[t]
    \centering
    \includegraphics[trim=0cm 0cm 0cm 0cm, clip,width=0.95 \columnwidth]{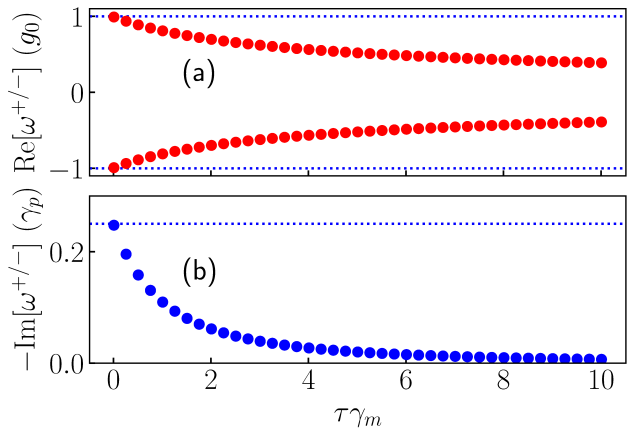}
    \caption{Complex poles of the two main polariton states 
    using Eq.~\eqref{eq:E3_exact} with no approximations. Note that $\tau=\tau_{\rm RT}/2$, and $\gamma_m=10\gamma_p$. The dashed lines show the non-retarded solution (Markov) as also identified in the text.} 
    %Note in these units, the results look identical with $\gamma_m=100\gamma_p$. }
    \label{fig:2}
\end{figure}

\begin{figure}[th]
    \centering
    \includegraphics[trim=0cm 0cm 0.3cm 0.0cm, clip,width=0.99 \columnwidth,width=0.95 \columnwidth]{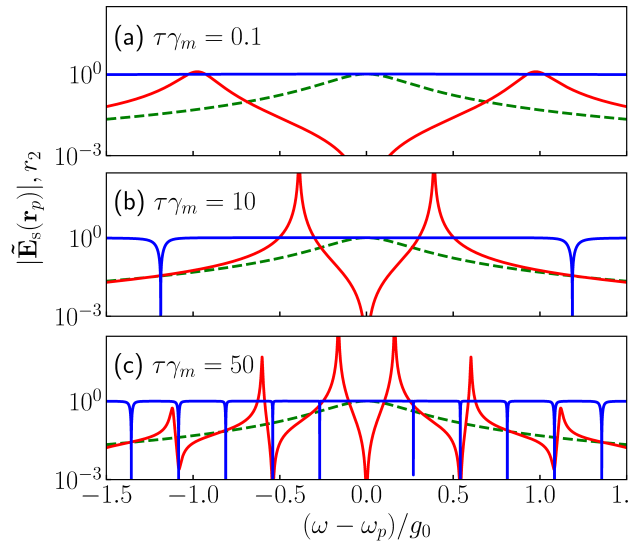}
    \caption{Scattered field (red curve) and two  qubit reflectivity (blue curve) at the probe dot for various delay times, using $\gamma_m=10\gamma_p$.
    (a) $\tau\gamma_m = 0.1$;
    (b) $\tau\gamma_m = 10$;
    (c) $\tau\gamma_m = 50$. The greed dashed curve shows the single qubit scattered field.}
    \label{fig:3}
\end{figure}

Figure~\ref{fig:2} shows the
first near-resonant complex poles as a function of
$\tau$ for a decay rate $\gamma_m = 10 \gamma_p$. In the Markovian limit, the value of the poles are constant (horizontal dashed lines). However, if the delay times are taken into account (non-Markovian case, symbols), it can be seen how the values of the complex poles decrease as the separation between the atoms increases; thus,  the resonances are in fact dependent on the retardation (and other modes can also become important, not shown here); however,  in  units of $g_0$, and using a Markov approximation, the coupling rate is always 1 (splitting of $\pm g_0$), but this only applies for sufficiently short delay times. When the retardation is \blue{sufficiently large}, we can have  $g_{\rm eff} < 1$ as $\tau$ increases, with already notable differences as small at $\tau\gamma_m=0.1$. In addition, we observe that in the units presented in Fig.~\ref{fig:2} ($\tau \gamma_m$), the results look identical for different  
decay rate ratios (e.g., if  $\gamma_m = 100 \gamma_p$).

In Fig.~\ref{fig:3}, we show the \blue{reflectivity} (blue curve) of the bare two-qubit mirror system
as well as the probe qubit scattered field (red curve) for different delay times. 
For the scattered field, we observe how Rabi splitting decreases and sharpens for longer delay times.
In contrast, the single qubit system just displays a Lorentzian lineshape (green curve).
In addition, for the longer delay \blue{time}, we can 
observe a coupling to the
\blue{$m=\pm2,4$} modes causing a striking Fano resonance~\cite{PhysRevA.95.053807,limonov_fano_2017};
note that the \blue{$m=\pm1,3$} modes do not couple
since the probe qubit is at a node of these cavity modes.
For the shorter delay time in 
  (a), the peaks are placed approximately at $\pm g_0$ (slightly decreased), which corresponds close to the \blue{Markovian} limit result. However,
  for the longer delay times, in (b) and (c), 
  then clearly the effective Rabi splitting, 
  $g_{\rm eff}$, becomes smaller for longer delay times.
 The first cavity sideband resonance in (b) and (c) are at
$\omega_{c1}=1.19g_0$ and $0.27g_0$, respectively;  and note that these are
also influenced by 
%not the same as the expected $\omega_{\rm FSR}$ because of
 the  dispersion in $r_1(\omega)$. Also note that  $\omega_{c1}/g_0 \propto \sqrt{\gamma_m/\gamma_p}$.

%\blue{\section{Quantum Theory using Matrix Products States (MPSs)}}
\textit{Quantum Theory using Matrix Products States (MPSs).---}To represent the time evolution of the probe qubit in the non-Markovian regime, and to confirm our semiclassical results in the frequency domain, we next solve the same system using MPSs \cite{yang_matrix_2018,vanderstraeten_tensor_2017,orus_practical_2014}. To do this, we first consider the Hamiltonian for three qubits \cite{PhysRevResearch.2.013238},
\begin{equation}
     H = \sum_{n=1,2,3} \omega_n \sigma^+_n \sigma^-_n
     + \sum_{\alpha=L,R} \int_{-\infty}^{\infty} d\omega  \omega b_\alpha^\dagger (\omega)b_\alpha(\omega)
     +  H_{\rm I},
%     \label{hamil3tls}
\end{equation}
where $\omega_n$ are the resonant frequencies and $\sigma^\pm_n$ are the Pauli operators of the qubits, and $b_\alpha(\omega)$ is the frequency dependent boson operator.
The interaction term, 
\begin{align}
    H_{\rm I} &=  \frac{1}{\sqrt{2\pi}}\int_{-\infty}^{\infty} d\omega\,
    \bigg \{\sum_{n=1,2,3} 
    \bigl( \sqrt{\gamma_{n}^L} e^{i\omega x_n/c}\, b_L(\omega)\sigma^+_n \nonumber \\
    &+  \sqrt{\gamma_{n}^R} e^{-i\omega x_n/c} b_R(\omega)\sigma^+_n \bigr) + \rm H.c. \bigg \},  
\end{align}
where $x_n$ are the positions of the atoms, and $\gamma_{n}^L$, $\gamma_{n}^R$ represent the left and right decay rates, respectively. 
\begin{figure}[t]
    \centering
    \includegraphics[trim=0cm 0.0cm 0cm 0cm, clip,width=1. \columnwidth]{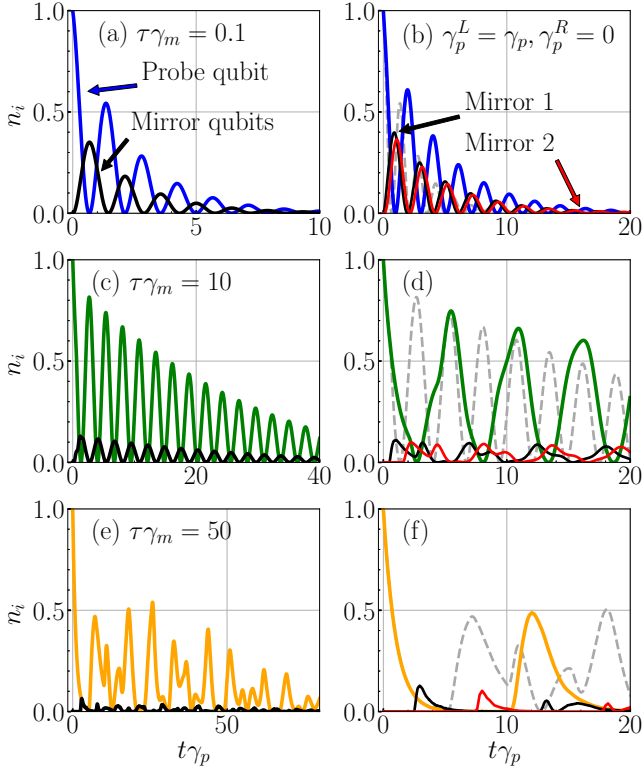}
    \vspace{-0.4cm}
    \caption{Probe and mirror population
    dynamics  using $n_p(0)=1$ in vacuum, where again we use  $\gamma_m=10\gamma_p$, for the same three delay times shown in Fig.~\ref{fig:3}:
    (blue) $\tau\gamma_m = 0.1$;
    (green) $\tau\gamma_m = 10$;
    (orange) $\tau\gamma_m = 50$.  
    Cases (a,c,e) are the symmetric examples corresponding to the ones studied in Fig.~\ref{fig:3} in the frequency domain, where $\gamma_p^L = \gamma_p^R = \gamma_p /2$. Subplots (b,d,f) show the same time dynamics when the probe qubit is chiral, with $\gamma_p^L = \gamma_p$ and $\gamma_p^R = 0$. The grey dashed line is the corresponding symmetric case for comparison.
    }
    \label{fig:4}
\end{figure}
% %
Choosing a rotating frame corresponding to the center (probe) qubit frequency, and defining the boson operators in the time domain, the Hamiltonian can be written as
\begin{align}
    &H =  \Delta_m \sigma^{+}_1 \sigma^{-}_1 + \Delta_m \sigma^{+}_3 \sigma^{-}_3 \nonumber \\
    &+ \sqrt{\gamma_{m}} \bigl(e^{i\omega_0 \tau} b_L(t{-}\tau) 
    +  b_R(t) \bigr) \sigma^+_1 + {\rm H.c.} \nonumber \\
    &+ \sqrt{\gamma_{p}} \bigl(  e^{i\omega_0 \tau /2} b_L(t{-}\tau /2) 
    +  e^{i\omega_0 \tau /2} b_R(t {-} \tau /2) \bigr) \sigma^+_2 + {\rm H.c.}  \nonumber \\
    &+ \sqrt{\gamma_{m}}  \bigl( b_L(t) 
    + e^{i\omega_0 \tau} b_R(t{-}\tau) 
    \bigr) \sigma^+_3 + {\rm H.c.},
    \label{timeHamil}
\end{align}
where we have considered symmetric decay rates, with $\gamma_1= \gamma_3 = \gamma_m$, $\gamma_2 = \gamma_p$, and the probe qubit is situated in the middle; $\Delta_m = \omega_m - \omega_0$ where $\omega_m$ is the frequency of the external qubits. For the cases considered, all qubits are on resonance and $\Delta_m = 0$.
Equation~\eqref{timeHamil} allows us to write the time evolution operator in the MPS representation as a matrix product operator  \cite{genericmpo}, 
\begin{align}
    U(t_k,t_{k+1}) &{=} \exp \bigl[ 
    -i \sqrt{\gamma_{m}} \bigl(e^{i\omega_0 \tau} \tilde B_L(t_{k-l}) 
    +  \tilde B_R(t_k) \bigr) \sigma^+_1  \nonumber \\
    &+ {\rm H.c.} -i \sqrt{\gamma_{p}} \bigl( \blue{ e^{i\omega_0 \tau/2}} \tilde B_L(t_{k-l/2}) \nonumber \\
    &+  e^{i\omega_0 \tau /2} \tilde B_R(t_{k-l/2}) \bigr) \sigma^+_2 + {\rm H.c.} \nonumber \\
    &-i \sqrt{\gamma_{m}}  \bigl( \tilde B_L(t_k) 
    + e^{i\omega_0 \tau} \tilde B_R(t_{k-l})
    \bigr],
\end{align}
where $\tilde B$ is the 
%boson 
time-bin bosonic
noise operator \cite{cajitas,PhysRevResearch.3.023030}.
%for the different time bins.

Following this procedure, the cases shown in Fig.~\ref{fig:3} are studied in the time domain and the evolution of the probe dot population ($n_p(t)=\braket{\sigma_2^+\sigma_2^-}(t)$) is represented in Figs.~\ref{fig:4} (a,c,e), \blue{and similarly for the mirror qubit populations}. Vacuum Rabi oscillations are present in all the cases. It is also observed that {\it significantly \blue{larger} decay rates appear for the probe dot for the longer
delay time} in (c) compared to (a), while the dynamics in (e) 
%where the decay rates $\gamma_m$ and $\gamma_p$ are the same. The last case presented 
($\tau \gamma_m = 50$) 
results in a high degree of interference because the qubit resonance now couples to the multiple cavity modes.  In addition, population evolution of the mirror qubits is shown in black (both mirror qubits show the \blue{same} population). We observe that this population stays closer to zero for longer delay times, and in (e) the interference is also reflected in the mirror dots.
%\blue{say something also about the mirror populations now}

%Finally, 
We next show further control of this 
waveguide cavity-QED system by
considering a chiral probe qubit with  $\gamma_p^L = \gamma_p$ and $\gamma_p^R = 0$. 
 Chiral systems allow  direction-dependent coupling, offering multiple applications in quantum networks (e.g., see Refs.~\cite{PhysRevLett.115.153901,PhysRevLett.117.240501,JalaliMehrabad:20}).
In Figs.~\ref{fig:4} (b,d,f), we show the time evolution for similar systems as before but now considering the chiral probe qubit. In this case, the photons only interact with the probe dot when they are reflected from the \blue{right} mirror. We also observe a variation of the time evolution with a further improvement of the coherence times. In addition, the mirror qubits results do not match as we have broken the symmetry of the system, and it can be seen that, when the delay times increase, the separation between the same peak for each mirror increases, showing very clearly the influence of the retardation on the chiral qubit dynamics.

\begin{figure}[t]
    \centering
    \includegraphics[trim=0cm 0.0cm 0cm 0cm, clip,width=1 \columnwidth]{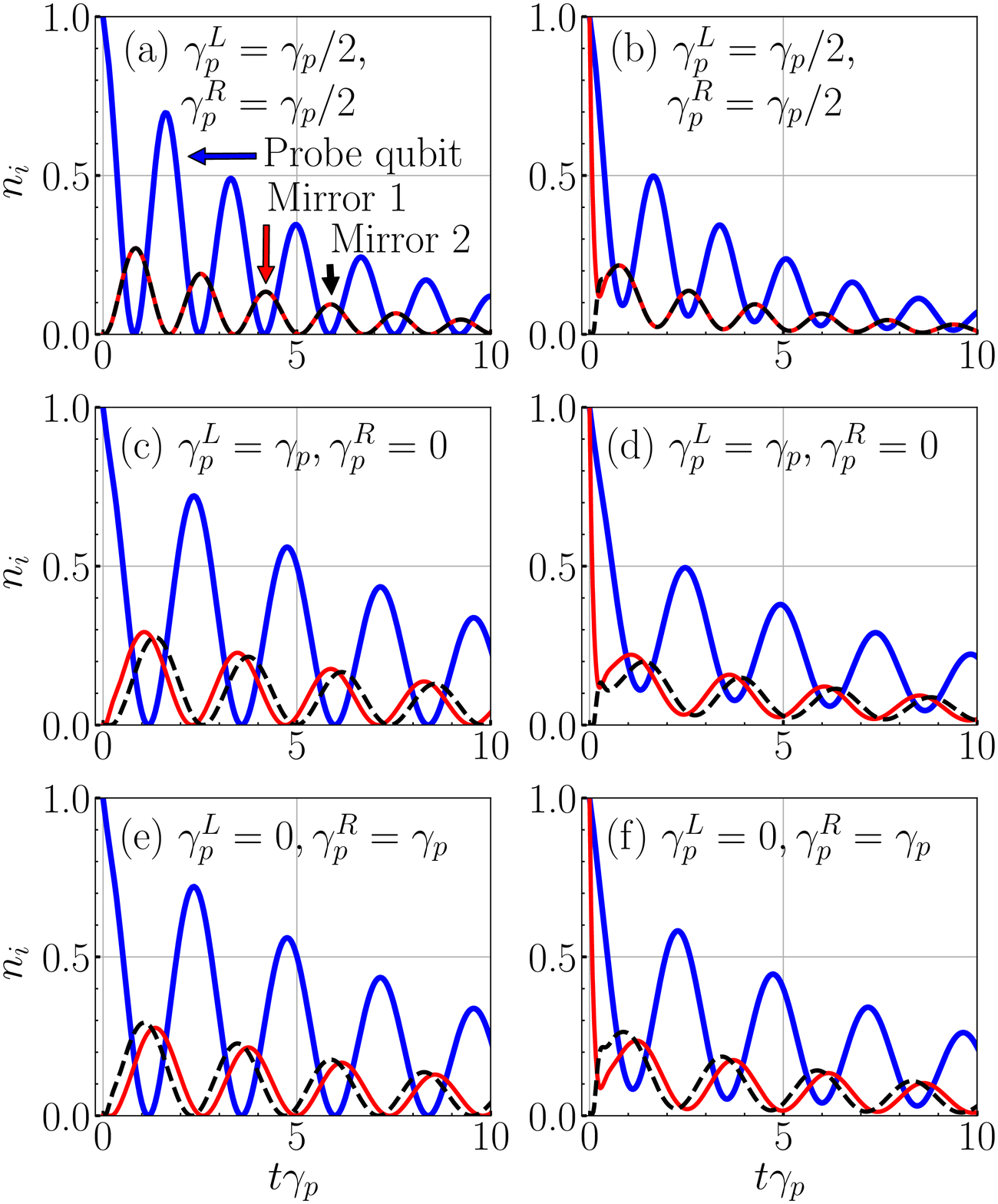}
    \vspace{-0.4cm}
    \caption{\blue{Probe and mirror population
    dynamics using one excitation ($n_p(0)=1$) in vacuum for (a,c,e), and two excitations ($n_p(0)=1$, $n_{m1}(0)=1$) in vacuum for (b,d,f). In all cases $\gamma_m=10\gamma_p$, for a delay time $\tau\gamma_m = 1$.
    Subplots (c,d,e,f) show the time dynamics when the probe qubit is chiral, with $\gamma_p^L = \gamma_p$ fand $\gamma_p^R = 0$ in (c,d), and $\gamma_p^L = 0$ and $\gamma_p^R = \gamma_p$ in (e,f). Although probe qubit behaves similarly in (c,e), the double excitation breaks the symmetry giving rise to different  quantum dynamics in (d,f).}
    }
    \label{fig:5}
\end{figure}
% %
%\vspace{1cm}

\blue{Finally, we consider a double excitation (nonlinear) case by also starting an excitation in one of the mirrors  (mirror 1 in Fig.~\ref{fig:5}). Here, we compare this new case with the one shown previously with just one excitation. For this example, we choose the  decay rate of $\gamma_m = 10\gamma_p$ and a feedback $\tau \gamma_m = 1$.
In Fig.~\ref{fig:5}(a,b), we show the symmetric coupling case ($\gamma_p^L = \gamma_p^R = \gamma_p /2$) and observe how the values of $n_p$ change but the oscillation period is similar. A significantly faster initial decay of the mirror qubit is seen in (b) due to the larger decay rate of the mirror qubits. Then we study the chiral cases, with $\gamma_p^L = \gamma_p$ and $ \gamma_p^R = 0$ in (c,d), and $\gamma_p^L = 0$ and $ \gamma_p^R = \gamma_p$ in (e,f). In (c,e), the mirror qubit are exchanged and the probe qubit behaves similarly. However, in the double excited cases (d,f), we can see that not only does $n_p$  vary but also the position of the peaks changes. In (d), only the photons coming from the right mirror (mirror 2) interact with the probe qubit, while in (f) the photons interacting with the probe qubit are the ones coming from the left one (mirror 1). This leads to higher values of $n_p$ in (f), and a phase difference between both scenarios.
}

%\blue{\section{Conclusions}}
\textit{Conclusions.---}We have introduced a theory of \blue{qubit-like} mirrors which 
\blue{mimics
 caviy-QED interactions,
 consistent with } recent experiments. 
We then extended this system 
into a much richer  dynamical coupling regime, 
 by introducing retardation through
a finite round trip memory, which increases the lifetime
of the cavity-qubit polariton states
by several orders of magnitude, while also reducing the overall vacuum Rabi splitting.
For sufficiently long delays, we  \blue{demonstrated} how several cavity modes can act in concern, producing non-trivial and highly
non-Markovian population dynamics of a probe qubit.
Our theory first presented an  analytical solution to the semiclassical problem of three dipoles in a waveguide, using an exact Green function.
% %without any rotating wave approximations. 
% An explicit solution was found for the scattered field 
% %and spectrum 
% from the probe qubit, from which a full set of complex poles can be analyzed.
% %, including the non-Markovian dressed polariton states associated with strong coupling. 
This picture was complemented by 
%solving the quantum dynamics 
using MPS, where we showed the  population  dynamics
including an example for an embedded
 chiral probe qubit. 
In both theoretical pictures, we explicitly showed how to increase the lifetime of the polariton states by increasing the delay time between the external qubit mirrors.
%, and showed that the Markovian dynamics is recovered for short delay times. 
\blue{In addition, we demonstrated the role of several quantum excitations in a nonlinear regime,
showing significantly faster early time decay.}
These retardation effects also
facilitate the exploration of   quantum nonlinear cavity-QED effects with \blue{coherent driving fields} using engineerable multiple qubits in waveguides.
%including
%chains of qubits.
%that can be controlled by mirror qubits 
% well into the deep non-Markovian coupling regime.

\vspace{0.2cm}
%\blue{\section{Acknowledgements}}
{\it Acknowledgements.---}We acknowledge funding from 
the Canadian Foundation for Innovation (CFI), Queen's University and
the Natural Sciences and Engineering Research Council of Canada (NSERC).

%\bibliography{refs,refs1}
\bibliography{biblio}

\end{document}